\documentclass[prl,twocolumn,tightenlines]{revtex4}%
\usepackage{amsfonts}
\usepackage{amsmath}
\usepackage{amssymb}
\usepackage{graphicx}%

\begin{document}
\title{Effective collective barrier for magnetic relaxation in frozen ferrofluids}%

\author{Ruslan Prozorov}
\email{prozorov@sc.edu}%
\affiliation{Department of Physics \& Astronomy and USC NanoCenter, University of South Carolina, 712 Main Street,
Columbia, SC 29208}%

\author{Tanya Prozorov}
\altaffiliation{Also at: School of Chemical Sciences, University of Illinois, Urbana, IL 61801}%
\affiliation{Department of Physics \& Astronomy and USC NanoCenter, University of South Carolina, 712 Main Street,
Columbia, SC 29208}

\author{Alexey Snezhko}
\affiliation{Department of Physics \& Astronomy and USC
NanoCenter, University of South Carolina, 712 Main Street,
Columbia, SC 29208} 


\begin{abstract}
Magnetic relaxation and frequency response were measured in frozen ferrimagnetic colloids of different concentrations.
A crossover from reversible to irreversible behavior is observed for concentrated colloids. In irreversible state,
magnetic relaxation is time-logarithmic over seven orders of magnitude of experimental time windows. A master curve
construction within mean field phenomenological model is applied to extract effective collective barrier as a function
of the irreversible magnetization. The barrier logarithmically diverges, providing evidence for self-organized critical
behavior during magnetic relaxation in frozen ferrofluids.
\end{abstract}

\received[Received: ]{2 December, 2003}

\maketitle

Slow magnetic dynamics in assemblies of superparamagnetic particles is a subject of intense experimental and
theoretical research. Frozen ferrofluids provide a model system for such studies due to controllable interparticle
separation and reduced agglomeration (compared to powders). Various models of single-particle
\cite{neel,stoner,jonsson1,lierop} and collective behavior \cite{jonsson1,luo,jonsson,zhang,mamiya,dormann} in these
systems were explored. It is generally agreed that collective effects play an important role, with long-range
dipole-dipole interactions as a dominant mechanism \cite{luo,dormann}. However, effects related to finite size of a
correlated volume for magnetic moment reorientation \cite{eberbeck,denisov} and possible spin-glass state
\cite{mamiya,dormann} add complexity to the problem. In all cases, the question of the effective energy barrier
governing magnetic relaxation and especially its dependence on the distance from the equilibrium is crucially
important. Knowing the evolution of the effective barrier with the driving force can help to distinguish between models
that predict its increase \cite{dormann} or decrease \cite{morup} with the increase of the strength of interparticle
interactions as well as sort out different relaxation regimes \cite{garanin}.

Phenomenological description of the magnetic relaxation provides a useful link between the experiment and the
microscopic models. The analogy of slow dynamics in the irreversible state of assemblies of magnetic nanoparticles with
similar phenomena of flux creep in type-II superconductors was first noticed in \cite{lottis} and later formulated in a
closed phenomenological form in \cite{prozorov}. The main advantage of this model is that it identifies the effective
barrier for magnetic relaxation as a single important parameter governing the process. Consider irreversible
magnetization $M=\left\vert M_{m}-M_{eq}\right\vert $, where $M_{m}$ is measured magnetization and $M_{eq}$ is the
equilibrium magnetization (i.e., obtained upon field cooling process). The microscopic models of the system dynamics
can be incorporated into a phenomenological equation \cite{prozorov}:
\begin{equation}
\frac{dM}{dt}=-\frac{M}{\tau_{0}}\exp\left[  -\frac{U(M)}{k_{B}T}\right]%
\label{dMdt}%
\end{equation}
\noindent where $\tau_{0}$ is the microscopic attempt time. The formulation of the problem in terms of the irreversible
magnetization as the driving parameter is experimentally verified by studying magnetic relaxation in various regimes of
inducing irreversible state (sweeping magnetic field, increasing magnetic field after zero-field cooling or decreasing
field after field cooling). The result is that regardless how the irreversible state was created, its dynamics can be
described by Eq.(\ref{dMdt}).

If the barrier $U(M)$ does not depend on irreversible magnetization, i.e., $U(M)=U_0$, the solution of Eq.(\ref{dMdt})
leads to a well-known N\'{e}el exponential relaxation:
\begin{equation}
M\left(  t\right)  =M_{0}\exp\left(  -\frac{t}{t_{N}}\right)%
\label{Neel}%
\end{equation}
\noindent where $t_{N}=\tau_{0}\exp\left(  U_{0}/k_{B}T\right)$ is the N\'{e}el relaxation time and $M_{0}$ is the
initial value of the irreversible magnetization. Eq.(\ref{Neel}) describes the dynamics of a uniform system of
non-interacting superparamagnetic particles \cite{neel}.

If, on the other hand, the barrier does depend on $M$, time - logarithmic magnetic relaxation is obtained
\cite{blatter}. For example, for a simplest linear dependence, $U=U_0+U_{c}\left(1-M/M_{c}\right)$, Eq.(\ref{dMdt})
yields:
\begin{equation}
M\left(t\right) =M_{c}\left(  1-\frac{k_{B}T}{U_{c}}\ln\left(  1+\frac{t}{t_{0}}\right)  \right)%
\label{Anderson}%
\end{equation}
\noindent where $t_0=\tau_Nk_BT/U_c$ is the effective time constant, and $M_c$ is the threshold value for the
irreversible magnetization above which the relaxation is of the N\'{e}el's exponential single-particle decay. Above
$M_c$, the barrier can still depend on $M$, but it would not lead to a collective dynamics and can be accounted for as
a renormalization of the Stoner-Wohlfart single-particle barrier \cite{stoner,garanin}. Time-logarithmic relaxation
described by Eq.(\ref{Anderson}) is routinely observed in assemblies of small ferromagnetic particles and, perhaps, is
the most controversial issue in this research subject. For a long time, the contra-argument has been that a wide
distribution of particle sizes and significant agglomeration cause distribution of the N\'{e}el relaxation times and
the resulting slow dynamics could be described by stretched exponents, leading to quasi-logarithmic relaxation. The
magnetization dependence of the effective barriers has also been analyzed in a single-particle approximation
\cite{jonsson1}. The single-particle models, however, cannot reproduce logarithmic dynamics in a wide range of
experimentally accessible times \cite{denisov}.
\begin{figure}[ptb]
\begin{center}
\includegraphics[width=8.0cm]{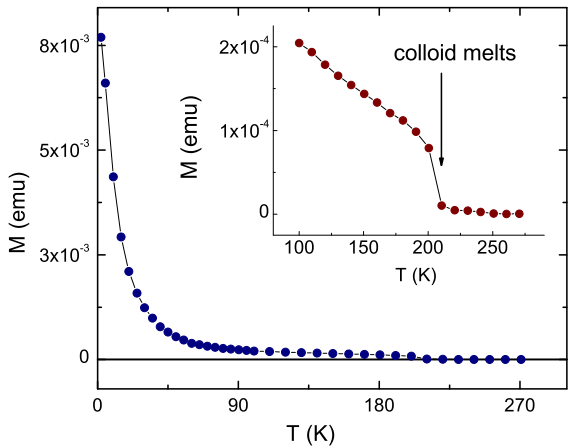}%
\caption{Measurements of remanent magnetization after field cooling in a $1$ T magnetic field from the liquid state and
removing the field at $5$ K with subsequent warming whence the measurements were performed. The inset shows zoomed high-temperature region.}%
\label{melting}%
\end{center}
\end{figure}

In this work, a set of non-oriented frozen ferrofluids containing spherical $\mathrm{Fe_2O_3}$ nanoparticles of 8-10 nm
in diameter, but different volume concentrations (interparticle distances) was studied by DC and AC magnetization
techniques. Time-logarithmic relaxation was observed in a wide experimental time domain, covering seven orders of
magnitude. This relaxation was observed at all temperatures and magnetic fields below the blocking line, $T_B(H)$, on a
$T-H$ phase diagram of the system. Within the framework of the adapted model \cite{prozorov}, the important question is
the dependence of the barrier for magnetic relaxation $U(M)$ on the irreversible magnetization, $M$. Typical slow
collective dynamics of interacting particles leads to a nonlinear and even diverging barrier on approaching the
equilibrium state. From a general point of view, such behavior is expected for a system evolving via many metastable
states, so that the effective time, $t_{e}=\tau_{0}\exp\left( U(M)/k_{B}T\right)$, is increasing due to a decrease of a
driving force (proportional to the irreversible magnetization, $M$). Various microscopic models predict different forms
of $U(M)$ - from strong algebraic divergence in a glassy phase \cite{eberbeck,blatter,yeshurun}, a weak logarithmic
divergence in the regime of self-organized criticality \cite{blatter}, to a non-diverging tilted-washboard barrier
\cite{blatter,yeshurun}. Therefore, the functional form of $U(M) $ is a key parameter, which links the macroscopically
observable dynamics to the microscopic mechanism of magnetic irreversibility and relaxation in assemblies of
ferromagnetic nanoparticles.

\begin{figure}[ptb]
\begin{center}
\includegraphics[width=8.0cm]{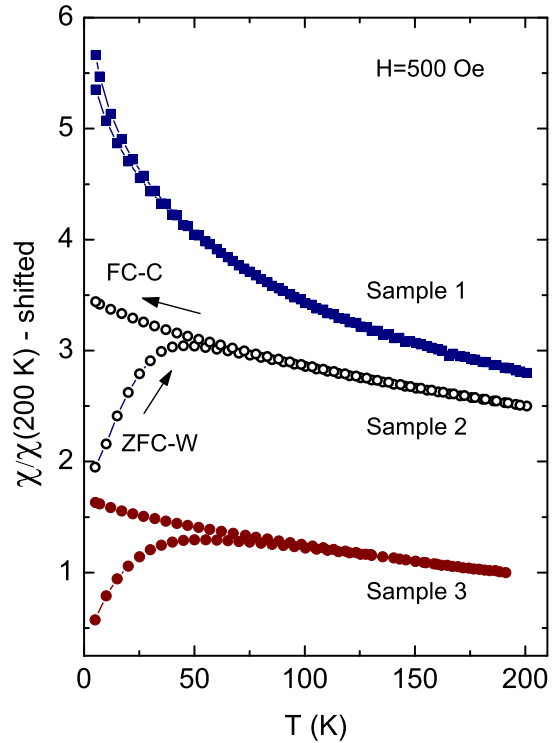}%
\caption{Zero-field and field-cooled magnetic susceptibility in three different colloids (from top to bottom, volume
fractions are $C_v \approx 0,2,0.3,0.5~\%$). The curves are offset for clarity.}%
\label{H500}%
\end{center}
\end{figure}

Colloids containing coated amorphous $\mathrm{Fe_2O_3}$ nanoparticles were prepared by a sonochemical method described
in details elsewhere \cite{suslick,shafi,prozorovt}. For all samples, the same amounts of oleic acid and decane were
used and the only variable was the amount of $\mathrm{Fe(CO)_5}$ added. Oleic acid acting as a surfactant, prevented
the particles from agglomeration. The size of obtained ferrimagnetic nanoparticles was about $8-10$ nm with a narrow
size distribution. Transmission electron microscopy (TEM) imaging showed no systematic difference between particle
sizes and shapes for colloids of different concentrations. Therefore, the only difference was the volume concentration
of $\mathrm{Fe_2O_3}$ nanoparticles. Samples studied in this work were true colloids - similar samples did not
precipitate even after about one year on the shelf, although the data presented here were collected on freshly obtained
materials. AC and DC magnetic measurements were performed on a 1 T \textit{Quantum Design} MPMS.

To verify the superparamagnetic nature of obtained colloids, the oriented frozen ferrofluid was measured. Figure
\ref{melting} shows the temperature dependence of remanent magnetization taken on warming from 5 K to room temperature.
Initially, the sample was slowly cooled down in a 1 T external magnetic field from room temperature to 5 K. This
allowed for a mechanical alignment of ferrimagnetic nanoparticles along the external field. At 5 K the external field
was removed and remanent magnetization was measured upon slow warming. Only at the melting temperature of a frozen
solution, the magnetization sharply dropped, indicating physical randomization of particles in the liquid phase. Each
particle behaves as a single magnetic moment of about $2000~\mu_B$ (Bohr magnetons) inducing significant long-range
dipole fields and therefore frozen ferrofluids represent a nearly ideal superparamagnetic system. Hereinafter, all
measurements are reported for zero-field cooled samples (random initial orientation of magnetic moments).

\begin{figure}[ptb]
\begin{center}
\includegraphics[width=8.0cm]{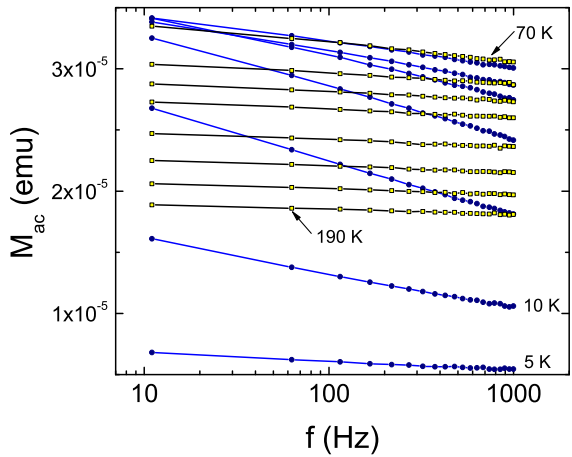}%
\caption{Frequency dependence of the AC magnetization measured at different temperatures for Sample 3 ($C_v \approx 0.5~\%$).
Solid symbols show curves below blocking temperature; open symbols - above.}%
\label{freq}%
\end{center}
\end{figure}

Figure \ref{H500} shows the development of magnetic irreversibility in frozen ferrofluids as the volume fraction
increases. Sample 1 had $C_v \approx 0.2~\%$ (volume percent), Sample 2 had $C_v \approx 0.3~\%$ and Sample 3 had $C_v
\approx 0.5~\%$. A magnetic field of 500 Oe was applied after zero-field cooling to 5 K, whence the measurements were
taken on warming (ZFC-W). Field cooling data were collected from 200 K (still below the melting point) without turning
off the magnetic field (FC-C). At very small concentration ($C_v < 0.1~\%$, not shown) the behavior is
superparamagnetic and fully reversible. Only a hint of developing irreversibility is observed in Sample 1 ($C_v \approx
0.2~\%$). For larger volume fractions, pronounced blocking and associated hysteresis is observed (Samples 2 and 3 in
Fig.~\ref{H500}). As expected, for larger concentrations, the blocking temperature is shifted to higher temperatures.
Although the observed development of irreversibility vs. volume fraction is in satisfactory agreement with earlier
reports \cite{dormann}, we point out that volume fraction is not a good parameter, because it is coupled to the
particle size and, therefore, materials with different interparticle distance can have the same $C_v$ and vice versa.
The more relevant parameter is the interparticle distance, $r$, which in our case of 10 nm particles was $r=50,45$ and
$40$ nm, for samples 1,2 and 3, respectively.

Figure \ref{freq} shows the amplitude of a linear AC response ($H_{ac}=1$ Oe) measured as a function of frequency at
different temperatures. At all frequencies, the maximum value is reached at a blocking temperature ($~40$ K in this
case) as usually reported \cite{jonsson1}. Clearly, the frequency dependence is logarithmic and can be described by an
equation similar to Eq.(\ref{Anderson}) with $t=1/f$. Therefore, Fig.~\ref{freq} provides evidence for
time-logarithmic relaxation in the interval between $10^{-3}$ and $10^{-1}$ seconds.%
\begin{figure}[ptb]
\begin{center}
\includegraphics[width=8.0cm]{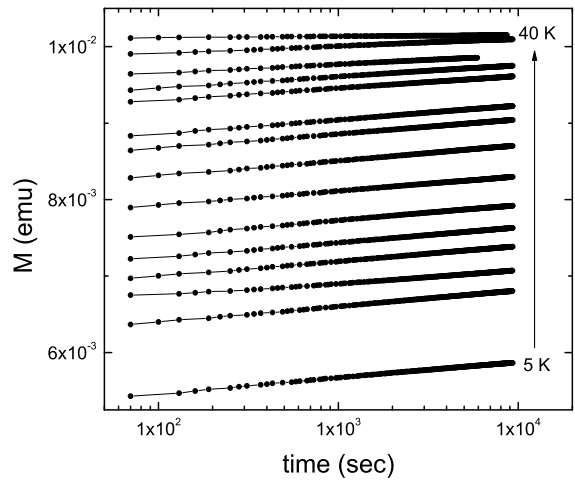}%
\caption{DC magnetic relaxation at different temperatures measured in Sample 2 ($C_v \approx 0.3~\%$)}%
\label{creep}%
\end{center}
\end{figure}
Figure \ref{creep} provides information on longer - times relaxation, probed by DC magnetization measurements. At all
temperatures, the relaxation is evidently time-logarithmic from 1 to $10^4$ seconds. Combined with the result of
Fig.~\ref{freq}, we obtain strong experimental evidence for a time-logarithmic magnetic relaxation in the irreversible
regime of frozen ferrofluids in the time interval of $10^{-3}-10^{4}$ seconds, which covers seven orders of magnitude.
It is very difficult to explain such a result by applying models of distributed barriers and stretched exponents.

Using collected relaxation data and Eq.(\ref{dMdt}) it is possible to estimate the dependence of the effective barrier
for magnetic relaxation on the irreversible magnetization. For this analysis, the construction, similar to that
developed for type-II superconductors \cite{maley} can be adapted. A somewhat related scaling approach was suggested
for ferrofluids in \cite{iglesias}. Rearranging Eq.(\ref{dMdt}) yields:
\begin{equation}
U(M)=k_BT\left(C-\ln{\left|\frac{d\ln{(M)}}{dt}\right|}\right)%
\label{U}%
\end{equation}
\noindent where $C=-\ln{\left(\tau_0\right)}$. The apparent problem is that the decay is logarithmic, so the logarithm
of the logarithmic relaxation rate changes insignificantly during any reasonable acquisition time. However, if those
relaxations are sampled at different temperatures as in Fig.~\ref{creep}, the wide range of relaxation rates,
corresponding to different values of the irreversible magnetization can be accessed.
\begin{figure}[ptb]
\begin{center}
\includegraphics[width=8.2cm]{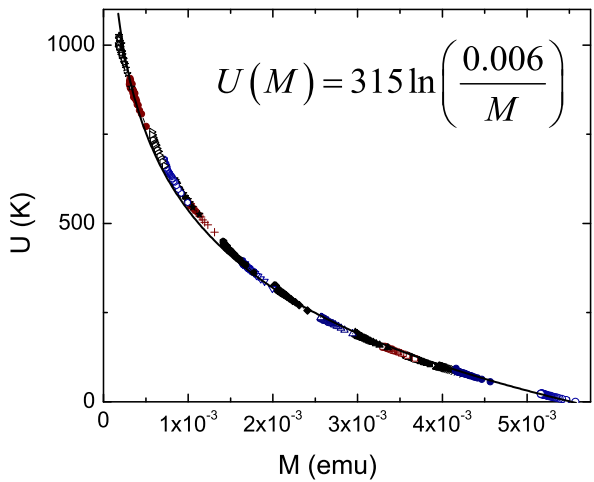}%
\caption{Effective collective barrier for magnetic relaxation constructed from
many measurements taken at different temperatures below blocking temperature (see Fig.~\ref{creep}).}%
\label{U}%
\end{center}
\end{figure}

Figure \ref{U} shows the corresponding experimental construction. Logarithmic derivatives of the individual $M(t)$
curves of Fig.~\ref{creep} were plotted against corresponding values of the irreversible magnetization and each curve
was adjusted to form a single master curve using $C$ as a free parameter. Required values of $C$ gradually increased in
agreement with the expected decrease of $\tau_0$. The analytical functional form of the obtained curve was examined by
fitting to various models, in particular to a general form of a collective barrier \cite{griessen},

\begin{equation}
U\left(  M\right)  =\frac{U_{c}}{n}\left(  1-\left(  \frac{M}{M_{c}}\right) ^{n}\right)%
\label{barrier}%
\end{equation}

\noindent which covers converging $\left(  n>0\right)  $, diverging $\left(  n<0\right) $ as well as logarithmic
$\left( n=0\right)$ barriers \cite{griessen}. It was found that the best fit is obtained for a logarithmic barrier,
$U\left( M\right) =U_{c}\ln\left( M_{c}/M\right)$, where $U_c\approx315$ K and $M_c\approx0.006$ emu. The logarithmic
barrier means that the effective relaxation time, $t_e$ depends algebraically on the irreversible magnetization,
$t_{e}=\tau_N\exp\left( U\left(  M\right)  /k_{B}T\right) =\tau_{0}\exp{U_0/k_BT}\left( M_c/M\right) ^{\gamma}$, where
$\gamma=U_{c}/k_{B}T$. The interpretation of the logarithmically diverging barrier is a subject of microscopic analysis
and it is hoped that this result experimental will stimulate further work. In superconductors, a logarithmic barrier is
obtained for two-dimensional interactions \cite{zeldov} as well as for magnetic relaxation in the regime of
self-organized criticality \cite{blatter,prozorog}. The latter is an interesting possibility for the relaxation in
frozen colloids, because self-organized criticality was already identified in random magnetic systems
\cite{onody,pazmandi,templeton}.

\begin{acknowledgments}
Acknowledgments: This work was supported by the NSF/EPSCoR under Grant No. EPS-0296165, a grant from the University of
South Carolina Research and Productive Scholarship Fund, and the donors of the American Chemical Society Petroleum
Research Fund. The TEM study was carried out in the Center for Microanalysis of Materials (UIUC), which is partially
supported by the DOE under Grant DEFGO2-91-ER45439.
\end{acknowledgments}

\end{document}